\documentclass[twoside,leqno,twocolumn]{article}

\usepackage[letterpaper]{geometry}

\usepackage{ltexpprt}
\usepackage{hyperref}

\usepackage{algorithm}
\usepackage{algorithmic}

\usepackage{amssymb}
\usepackage[utf8]{inputenc} 
\usepackage{url}            
\usepackage{booktabs}       
\usepackage{amsfonts}       
\usepackage{nicefrac}       
\usepackage{microtype}      
\usepackage{amsmath}
\usepackage{multirow}
\usepackage{tikz}
\usetikzlibrary{calc}
\usepackage{xcolor}         
\usepackage{makecell}       
\usepackage{subcaption}
\definecolor{empcolor}{gray}{0.90}

\newcommand*{\boxcolor}{gray}
\makeatletter
\renewcommand{\boxed}[1]{\textcolor{\boxcolor}{%
\tikz[baseline={([yshift=-0.5ex]current bounding box.center)}] \node [rectangle, minimum width=1ex,rounded corners,draw] {\normalcolor\m@th$\displaystyle#1$};}}
 \makeatother

\newcommand\ChangeRT[1]{\noalign{\hrule height #1}}
\newcommand{\algo}{{\textsf{SIM}}}
\definecolor{empcolor}{gray}{0.90}
\newcommand\outl[1]{\colorbox{empcolor}{#1}}
\DeclareMathOperator{\var}{\mathbb{V}}
\DeclareMathOperator{\mean}{\mathbb{E}}
\DeclareMathOperator*{\argmax}{argmax}

\begin{document}

\newcommand\relatedversion{}


\title{\Large Understanding Influence Maximization via Higher-Order Decomposition}
\author{Zonghan Zhang
\and Zhiqian Chen}

\date{Department of Computer Science and Engineering, Mississippi State University\\
     zz239@msstate.edu, zchen@cse.msstate.edu}

\maketitle


\fancyfoot[R]{\scriptsize{Copyright \textcopyright\ 2023 by SIAM\\
Unauthorized reproduction of this article is prohibited}}





\begin{abstract} \small\baselineskip=9pt 
Given its vast application on online social networks, Influence Maximization (IM) has garnered considerable attention over the last couple of decades. 
Due to the intricacy of IM, most current research concentrates on estimating the first-order contribution of the nodes to select a seed set, disregarding the higher-order interplay between different seeds.
Consequently, the actual influence spread frequently deviates from expectations, and it remains unclear how the seed set quantitatively contributes to this deviation. 
To address this deficiency, this work dissects the influence exerted on individual seeds and their higher-order interactions utilizing the Sobol index, a variance-based sensitivity analysis. To adapt to IM contexts, seed selection is phrased as binary variables and split into distributions of varying orders.
Based on our analysis with various Sobol indices, an IM algorithm dubbed~\algo~is proposed to improve the performance of current IM algorithms by over-selecting nodes followed by strategic pruning.
A case study is carried out to demonstrate that the explanation of the impact effect can dependably identify the key higher-order interactions among seeds.~\algo~is empirically proved to be superior in effectiveness and competitive in efficiency by experiments on synthetic and real-world graphs.
\end{abstract}

\section{Introduction}

As online social networks have drawn and maintained hundreds of millions of users during the past few decades, influence maximization (IM) attracts great attention~\cite{domingos2001mining,li2018influence}. It is widely applied to viral marketing~\cite{chen2010scalable}, rumor control~\cite{he2012influence}, social recommendation~\cite{ye2012exploring}, and infectious disease containment~\cite{newman2002spread}. 
IM is an NP-hard task in which a $k$-sized seed set is selected to maximize the number of influenced nodes which is also called influence spread~\cite{kempe2003maximizing}. 
The ineffectiveness or unreliability of IM methods could result in the loss of millions of dollars or even human lives, so their performance and dependability are of the utmost importance.

Current IM research suffers two major flaws. 
\textbf{(1) Lack of tools for influence spread decomposition.}
Most current research focuses on identifying a set of seeds that maximizes the expected influence spread. Other than that, which seed or seeds account for the greatest contribution to the final influence has not been discussed. Although some of the heuristics select seeds according to transparent measures such as centrality metrics, the process through which the seed set results in the final influence spread is still hidden. Nonetheless, due to the high stake in the quality of the solution, this explanation of the selected seed set is necessary. Meanwhile, existing tools that can disentangle and quantitatively evaluate the contributions of the input variables in a nonlinear function have various drawbacks:
\textit{Local interpretation methods}, including LIME~\cite{ribeiro2016should} and Grad-cam~\cite{selvaraju2017grad}, focus only on local neighborhoods and disregard the global impact of the seeds and their higher-order interactions.
\textit{Marginal contribution}~\cite{catav2021marginal} does not separate the higher-order interaction effects from the main effect thus fails to quantify the interactions among seeds.
\textit{The Shapley value}~\cite{roth1988shapley} distributes the higher-order interaction among the seeds and tends to underestimate the contribution of the nodes with larger influence overlaps with others~\cite{owen2014sobol}.
Beyond that, none handle higher-order interactions in the IM scenario well.
\begin{figure}[htpb!]
  \centering
  \includegraphics[width=0.85\linewidth]{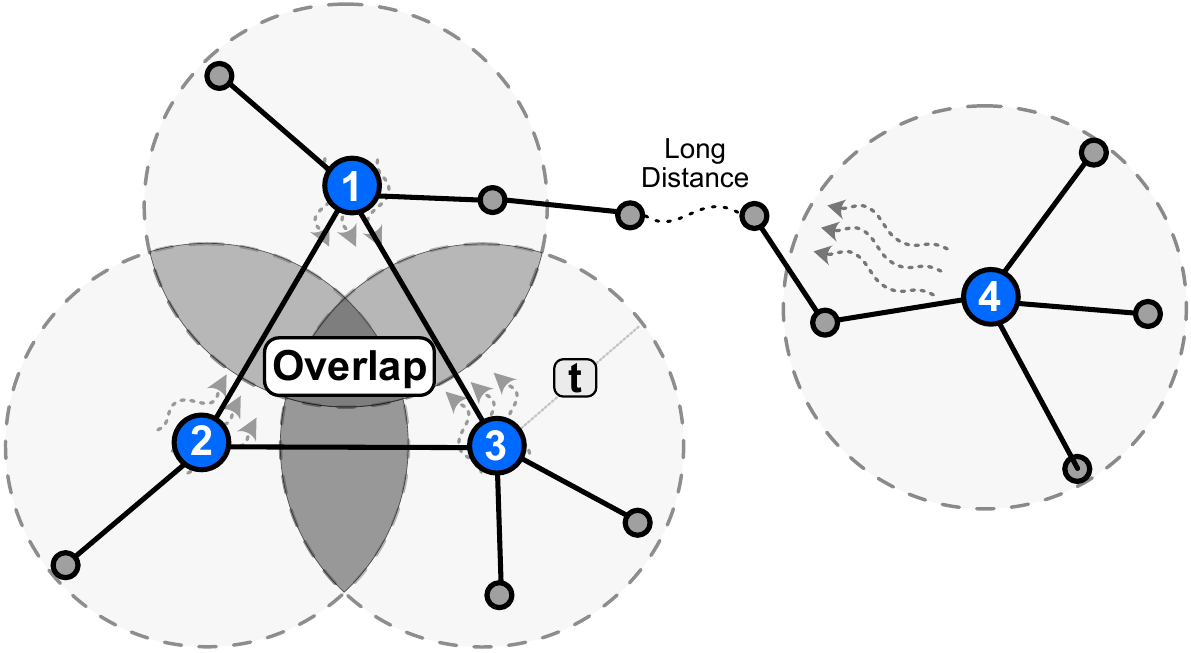}
  \caption{Interaction among the seeds.}
  \label{fig:1}
\end{figure}

\noindent\textbf{(2) The imbalance of effectiveness and efficiency.}
Current seed selection methods are typically divided into two categories, simulation-based and proxy-based~\cite{li2018influence}. Unfortunately, neither can offer consistently trustworthy solutions in a timely manner. 
\textit{The simulation-based methods} are theoretically guaranteed to have a high approximation performance by greedily adding the node with the highest marginal spread in each step~\cite{kempe2003maximizing}. 
However, multiple rounds of simulations are required to evaluate the marginal spread for each candidate node in each round, making the heuristic extremely slow when the graph size is large~\cite{arora2017debunking}. 
For the sake of efficiency, \textit{the proxy-based IM algorithms} 
emphasize the first-order contribution while disregarding the interaction efficacy among the seeds~\cite{chen2009efficient, yan2019minimizing, zhang2022blocking}. 
Nevertheless, due to the combinatorial impact in the seed set, the actual influence spread is heavily influenced by this interaction~\cite{kempe2003maximizing, chen2009efficient, li2018influence}. Thus, there is no assurance that the proxy-based algorithms can reach a good approximation ratio.
Figure~\ref{fig:1} illustrates an example of this interaction by depicting the influence distribution of a four-node seed set. Three of them (nodes 1, 2, and 3) are close to one another, whereas the fourth is far away. During $t$ stages of the propagation, the influence flows propagated from the first three seeds may overlap (gray areas), canceling a portion of their overall influence spread and failing to maximize the joint coverage. 
In IM community, most research still focuses on searching for faster but less accurate simulation methods~\cite{cheng2013staticgreedy, borgs2014maximizing} or designing better proxies that usually come with higher computational loads~\cite{yan2019minimizing, zhang2022blocking}. Still, the gap between effectiveness and efficiency is yet to be filled.

To address these issues, we introduce a global sensitivity analysis method, namely the Sobol indices~\cite{sobol1993sensitivity, wagener2019has}, as an effective method to decompose a seed set's influence spread.
In particular, the global contribution of a seed node is measured using the Sobol total index, which reflects its overall contribution to the influence spread.
The contribution of the node is then explained using first- and higher-order Sobol indices. Notably, the interactions between seed nodes are identified quantitatively with a higher-order index, which explains why most first-order approaches fail to identify the contributions of nodes appropriately.
The Sobol indices can overcome the shortcomings of the alternative tools mentioned above: 
\textbf{(1)} It is a global sensitivity method that considers a seed's global impact within the seed set. 
\textbf{(2)} The first-order effect and the higher-order interactions can be separated with a clear cut utilizing the corresponding Sobol indices.
\textbf{(3)} 
A seed's Sobol total index includes the complete contribution of all of its higher-order interactions with the rest of the seed set. Hence it never overestimates the influence like Shapley value does~\cite{owen2014sobol}.

After decomposing the influence spread towards each node, we rank the nodes according to their relative importance defined by their contributions. With this information, we propose a light-weighted method named~\algo~to find a balance between effectiveness and efficiency. Specifically,~\algo~selects excessive candidates and uses the Sobol indices to identify the ones that need to be excluded.
Our primary contributions include:
\begin{itemize}
\item \textbf{Explain and decompose the influence spread}: 
Given any seed set and the corresponding graph, each seed's contribution to the influence spread can be measured with the proposed explanatory method regardless of the IM algorithm that generates the seed set. Thus, this method can serve as a universal post-hoc explanation of an IM algorithm.
To the best of our knowledge, this paper serves as the first discussion on quantifying the influence overlap among the seeds. 
\item \textbf{Provide a new IM schema}:
By combining the simulation with the proxies,~\algo~overcomes the scalability and dependability issues in the literature. It suggests a new direction for IM research and provides a new schema for designing IM methods. 
\item \textbf{Conduct extensive empirical experiments}:
we prove that the Sobol indices can reliably evaluate the contribution of the nodes in the seed set with a case study.
Real-world and synthetic datasets are employed to demonstrate that~\algo~achieves superior performance in a reasonable amount of time.

\end{itemize}

\section{Related Work}

\textbf{Influence Maximization}.
Given that IM is NP-hard, researchers attempt to find a feasible solution with the best performance. As the first approximation attempt, a simulation-based greedy algorithm was proposed in~\cite{kempe2003maximizing}, which is not scalable. Similarly, a thread of simulation-based methods is developed to increase the performance or reduce the complexity~\cite{leskovec2007cost, goyal2011celf++}. Although great effort has been made to accelerate the process of the simulation-based methods, the complexity is still unacceptably high for the enormous online social networks~\cite{arora2017debunking, tang2014influence}. Most importantly, the opaqueness of the simulation makes this thread of methods impossible to be explained or improved by evaluating the diffusion process~\cite{li2018influence}. To alleviate the heavy computational burden of simulations, researchers turned to proxy-based methods in which the spreading power of the nodes is estimated by specific proxies. They started from simple heuristic measures such as degree, PageRank~\cite{page1999pagerank}, eigen-centrality~\cite{zhong2018identifying} and later turned to influence-aware or diffusion model-aware proxies~\cite{chen2009efficient, kimura2009blocking, yan2019minimizing,zhang2022blocking} to better estimate the influence spread brought by the seeds.

\noindent\textbf{Variance-Based Sensitivity Analysis}. Sensitive analysis studies the proportion based on which the uncertainty lying within the output of a model is distributed to the multiple sources of uncertainty in the input variables~\cite{saltelli1995use}. Variance-based methods take a major part in sensitive analysis and have been dated back to 1970s~\cite{cukier1973study}. Since then, these methods have been well studied and widely adopted by researchers and practitioners~\cite{saltelli2008global}. Among them, the Sobol indices~\cite{sobol1993sensitivity} is considered a significant milestone. However, variance-based methods cannot be directly utilized on IM problems since the input variables are not clearly defined and the influence spread has not been represented as a nonlinear function. In the following section, we will adapt the Sobol indices and a few of its important inheritors to the IM context and utilize them to provide a human-understandable explanation for a given seed set.

\section{Problem Setup}

A network is denoted by a bi-directional graph $G = (\mathcal{V}, \mathcal{E})$ where $\mathcal{V}$ and $\mathcal{E}$ represent vertices and edges respectively. 
Given the graph $G$, a seed budget $k\in \mathbb{N}^{+}$, and an influence maximization method $M$, a k-size seed set $\Omega = M(G, k)$ can be generated to approximately maximize the expected influence spread. 

(1) Our first objective is to develop \textit{\textbf{an explanatory method}} that can explain the total contribution of a single candidate seed node from a higher-order perspective.
The $\underline{\textbf{T}}$otal contribution $f_i^{(T)}$ of seed $i$ towards the expected influence spread $\sigma(\Omega)$ can be decomposed to different orders of $i$ such that: 
\begin{equation}\small
    f_i^{(T)} = \sum_{\Psi\subseteq \Omega} f_\Psi,
\end{equation}
where $\Psi$ is all the possible subsets containing node $i$, and $f_\Psi$ is the contribution of the interaction involving all nodes within a seed set $\Psi$. 
Suppose $\Omega = \{i, j, k\}$, then $\Psi \in \{\{i\},\{i, j\}, \{i, k\}, \{j, k\},\{i, j, k\}\}$. Note that $\Psi=\{i\}$ means the first-order contribution.
$f_\Psi$ is the contribution of the highest-order interaction in $\Psi$. For example, when $\Psi = \{i, j, k\}$, $f_\Psi$ represent the interaction among $i, j, k$ together, with no pairwise interaction such as the interaction between $i$ and $j$. 

(2) Our second objective is to design \textit{\textbf{an efficient and effective IM schema}} to identify the most influential seed set ($\Omega^{*}$) of size $k$ in terms of higher-order interactions, such that:
\begin{equation}\small
    \Omega^{*} = \argmax_\Omega (\sum_i f_{\{\Omega_i\}} + \sum_i \sum_{j>i} f_{\{\Omega_i,\Omega_j\}} + \dots + f_{\Omega})
\end{equation}
in which $\Omega_i$ is the $i$-th node of $\Omega$.

\section{Preliminary: Functional Analysis of Variance} \label{sec:prelim}
The objective of the functional analysis of variance (ANOVA) is to assess the significance of input to output based on the variance relationship between them. 
In this section, we will introduce ANOVA with \textit{first-order effect}, \textit{higher-order effects} and \textit{total effect}, which is defined by Sobol indices~\cite{SALTELLI2010259} and represent significance from different perspectives.
Specifically, variance-based functional decomposition is employed to analyze the impact of each variable.
Given any model of the form $Y=f\left(X_{1}, X_{2}, \ldots X_{n}\right)$, with $Y$ a scalar, the steps of a variance-based framework are as follows:
\begin{equation}\small
    Y=f_{0}+\overbrace{\sum_{i} f_{i}}^{\text{first-order}}+\overbrace{\sum_{i} \sum_{j>i} f_{i j}\quad+\cdots+\quad f_{12 \ldots n}}^{\text{higher-order}},
\end{equation}where
{\small
\begin{align*}
& f_{0}=\mean(Y), \\
& f_{i}=\mean_{\mathbf{X}_{\sim i}}\left(Y \mid X_{i}\right)-\mean(Y), \\
& f_{i j}=\mean_{\mathbf{X}_{\sim i j}}\left(Y \mid X_{i}, X_{j}\right)-f_{i}-f_{j}-\mean(Y),
\end{align*}}and similarly for the higher-order terms. $X_{i}$ is the $i$-th variable, and $\mathbf{X}_{\sim i}$ denotes the variable set $\mathbf{X}=\{X_1,X_2, \ldots, X_n\}$  that excludes $X_{i}$. $\mean$ denotes the expectation.
So $f_i$ can be explained as the expected value change of $Y$ taken over $\mathbf{X}_{\sim i}$ before and after $X_i$ is fixed, deriving the contribution of $X_i$ to $Y$.
Similarly, $f_{ij}$ means the expected value change of $Y$ before and after fixing $X_i$ and $X_j$. 
We will demonstrate below how to calculate the variance of a variable in various orders.

\noindent\textbf{(1) First-Order Effect}, denoted as $\mathbf{V}_i$, is the impact through one single variable only, which is
 represented by variance $\var$ accounted for $X_{i}$ and it can be written as:
{\small\begin{equation}
   \mathbf{V}_{i}:= \var\left(f_{i}\right)=\var_{X_{i}}[\overbrace{\outl{$ \mean_{\mathbf{X}_{\sim i}}\left(Y \mid X_{i}\right)$}}^{\text{inner expectation}}],
   \label{eq:1st_analysis}
\end{equation}}since $\var_{X_{i}}[\mean(Y)]=0$. $\var_{X_{i}}(\cdot)$ represents the variance of argument ($\cdot$) taken over $X_{i}$ .The meaning of the \textit{inner expectation} is that the mean of $Y$ is taken over all possible values of $\mathbf{X}_{\sim i}$ while keeping $X_{i}$ fixed. The outer variance ($\var_{X_{i}}$) is taken over all possible values of $X_{i}$. 
The associated first-order sensitivity is obtained by normalization: 
{\small\begin{equation}
    S_{i}=\frac{\mathbf{V_i}}{\var(Y)}.
    \label{eq:1st_sobol}
\end{equation}}





\noindent\textbf{(2) Higher-Order Effect}, using a second-order interaction effect between $X_i$ and $X_j$ as an example, can be expressed as:
{\small\begin{equation}
\mathbf{V}_{i j}^{(H)}:=\var[f_{ij}(X_i, X_j)] = \mathbf{V}_{ij}-\mathbf{V}_{i} - \mathbf{V}_{j},
\label{eq:higher_var}
\end{equation}}which means the variance brought by the interactions between variables $i$ and $j$. $\mathbf{V}_{ij}$ means the first-order variance of the variable set $\{i, j\}$, and it considers the variance of $\{i\}, \{j\}, \text{or} \{i,j\}$. Extending Equation \ref{eq:1st_analysis} from an individual variable to a variable set, we have:
\begin{equation}
    \mathbf{V}_{ij}=\var_{X_{i} X_{j}}\left[\mean_{\mathbf{X}_{\sim i j}}\left(Y \mid X_{i}, X_{j}\right)\right].
\end{equation}
Further, Equation \ref{eq:higher_var} can be normalized by $\var(Y)$:
\begin{equation}
\small
    S^{(H)}_{ij} = \frac{\mathbf{V}_{i j}^{(H)}}{\var(Y)} = S_{ij} - S_i - S_j,
    \label{eq:higher_sobol}
\end{equation}
which means that the second-order effect is the first-order effect of the variable set minus the first-order effects of both individual variables.
Similarly, we can calculate the third-order interaction among variables $X_i, X_j,$ and $X_h$ as
\begin{equation}
    S^{(H)}_{hij} = S_{hij}  - S_h - S_i - S_j-  S^{(H)}_{hi} - S^{(H)}_{hj} - S^{(H)}_{ij},
    \label{eq:higher_sobol_three}
\end{equation} and so on. Given there are totally $n$ variables in the model, we have
\begin{equation}\small
    \sum_i S_i + \sum_i \sum_{j>i} S^{(H)}_{ij} + \dots + S^{(H)}_{12\dots n} = 1.
\end{equation}

\noindent\textbf{(3) Total Effect} is the impact through one variable and all the interactions with the other variables, i.e., first- and higher-order effects. It can be calculated with \cite{homma1996importance}:
{\small\begin{equation}
    S^{(T)}_{i}=\frac{\mean_{\mathbf{X}_{\sim i}}\left(\var_{X_{i}}\left(Y \mid \mathbf{X}_{\sim i}\right)\right)}{\var(Y)},
    \label{eq:total_sobol}
\end{equation}}where $S^{(T)}_{i}$ measures the total effect, i.e. sum of all first- and higher-order effects involving $X_{i}$. 

\section{Higher-Order Influence Decomposition}

We will use the Sobol indices described in the preliminary to examine the current IM algorithms. Specifically, we will discuss how to (1) \textbf{explain} how higher-order effect characterizes the contribution of which first-order methods fail to do; and (2) \textbf{estimate} the contribution of node selection with total effect.

As discussed in the last section, sensitivity analysis is based on continuous variables of a nonlinear function. Nevertheless, the node selection problem in IM has not been described as such a function before.
To bridge the gap, we define node selection as binary variables in the IM problem, representing whether a seed is included in the optimal seed set since a node can be either selected or not while it can not be partially selected.
Specifically, a uniform distribution is employed such that each variable can be either $0$ or $1$ with a $50\%/50\%$ probability. 
Similar to Sobol notation, $\Omega_i$ denotes the selection decision for the $i$-th seed (1 for selected, 0 for not). 
This adaptation would greatly simplify the estimation of the Sobol indices since the variables in the model are bounded to be binary, limiting the number of combinations to $2^k$. 
Comparatively, the time complexity of calculating the Sobol indices of a model including not only binary variables is very high. For discrete variables, the number of combinations increases exponentially along with the number of categories. And continuous variables can be considered as categorical variables that have infinite categories.

With the binary variables representing the selection decisions of the nodes, the randomness in the model is brought by the uncertainty in the influence process. In our research, we adopt both the IC model and the LT model~\cite{kempe2003maximizing} to mimic the influence process. The propagation spreads in multiple discrete time steps $t_i$. At $t_0$, some initial nodes are active as seeds. 
The uncertainty is brought by the propagation probability for the IC model and the random assignment of the node thresholds in each round of simulation for the LT model. Details are discussed in Appendix.

\subsection{Explain with First- and Higher-order Effects.}
To explain the significance of the seed, its influence spread is decomposed into first- and higher-order effects. We will illustrate how to calculate first-order and higher-order impacts in an IM setting and identify which higher-order relationship contributes so substantially that a first-order analysis alone is insufficient.

\noindent\textbf{(1) First-Order Effect in IM.}
The first-order Sobol index of the node $i$ evaluates the influence spread it brings with no overlaps with other nodes. This is the area activated by the node $i$ as the seed, while no other seeds could successfully activate it. Representing the selection of the node $i$ with a binary variable, we can calculate the first-order Sobol index inspired by Equation~\ref{eq:1st_sobol}:
\begin{equation}
\small
\begin{split}
S_i &= \frac{\var_{\Omega_i}(\mean_{\Omega_{\sim i}}(Y|\Omega_i))}{\var(Y)}  
    = \frac{1}{4^k \cdot \var(Y)}[\sum_{\Omega_{\sim i}} Y_{\Omega_{i}=1} - Y_{\Omega_{i}=0} ]^2,
\end{split}
\label{eq:1st_im}
\end{equation} where $\Omega_{\sim i}$ represents a subset of the seed set that excludes the $i$-th seed.
See Appendix for transformation details of Equation \ref{eq:1st_im}.
Define $\Delta_{\Omega_{\sim i}} :=| Y_{\Omega_i = 1} -Y_{\Omega_i = 0}|$ as the difference of influence spreads of $\Omega_i$ given seed sets $\Omega_{\sim i}$, we can rewrite Equations~\ref{eq:1st_im} as:
\begin{equation}
    S_i = \frac{(\sum_{\Omega_{\sim i}} \Delta_{\Omega_{\sim i}})^2}{4^k\cdot \var(Y)},
    \label{eq:short_1st_im}
\end{equation}which will be used to quantify the first-order effect of certain nodes and to compare the higher-order effects.

\begin{figure}[htpb!]
  \centering
  \includegraphics[width=0.85\linewidth]{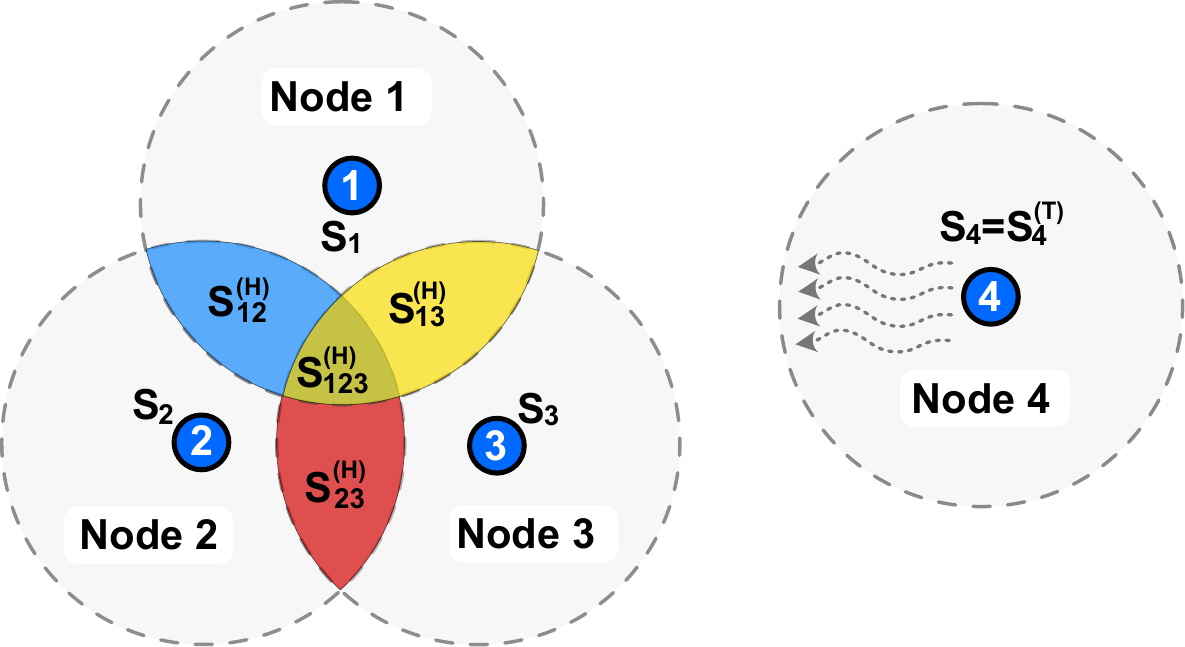}
  \caption{Sobol-based Influence decomposition.}
  \label{fig:2}
\end{figure}

\noindent\textbf{(2) Higher-Order Effect in IM.}
In the IM scenario, the higher-order Sobol indices can quantify the influence overlaps among the nodes and help with identifying the critical overlaps that make the IM algorithms misjudge the value of selecting a certain node as one of the seeds. The larger a node's total higher-order Sobol indices are, the more the actual influence spread it results in would shrink from its value identified by the IM algorithms. These indices can be calculated either by summing the higher-order Sobol indices from all orders together or by subtracting the first-order Sobol index from the Sobol total index, 
The calculation of the higher-order Sobol indices can also be adapted to the IM settings and simplified. Based on~\ref{eq:1st_analysis}, the first-order Sobol index of a size-$s$ subset $\Psi \subseteq \Omega$ can be calculated by:
\begin{equation}\label{eq:Sho}
\small
\begin{aligned}
    S_\Psi &= \frac{\var_\Psi  (\mean_{\Omega_{\sim \Psi}}(Y|\Psi))}{\var(Y)} 
    =  \frac{\sum_\Psi (\sum_{\Omega_{\sim \Psi}} (Y|\Psi) \cdot 2^{s-k} - \mean(Y))^2}{2^s\cdot\var(Y)}.
\end{aligned}
\end{equation}
See Appendix for transformation details of Equation \ref{eq:Sho}.
The higher-order Sobol indices can be calculated in an iterative manner, starting from the second order to the $k$-th order. Specifically, the $s$-th order Sobol indices of $\Psi$ whose $|\Psi| = s$ is: 
\begin{equation}\small
    S_{\Psi}^{(H)} = S_{\Psi} - \sum_{i=1}^{s} S_i - \sum_{|\zeta|=2}^{|\zeta|=(s-1)}\sum_{\zeta} S^{(H)}_{\zeta},
\label{eq:higher_sobol_many}
\end{equation}where $\zeta\subseteq\Psi$. Note that there exist multiple subsets for each order. 
For example, suppose $\Psi=\{1,2,3\}$ as in Figure~\ref{fig:2}, the implementation of Equation \ref{eq:higher_sobol_many} is:
\begin{equation}
    S^{(H)}_{123} = S_{123} - S_1 - S_2 - S_3 - S^{(H)}_{12} - S^{(H)}_{23} - S^{(H)}_{13}
\end{equation}
Similarly, it is possible to compute the fourth-order and even higher-order effects. 
Intuitively, though, higher-order relationships have fewer effects on average, such as $S^{(H)}_{123}$ being smaller than $S^{(H)}_{12}$, yet incur exponentially rising computing costs. 
In light of this, this study will investigate the second-order contributions to influence propagation, which, as proved by our experiments, are sufficient to explain why first-order is insufficient. Various higher-order selections can be expanded as necessary.
\subsection{Estimate influence spread of Node Set with Total Effect.} 
Given a seed set $\Omega$ and the graph $G$ representing the network where the propagation happens, we can estimate the influence spread within $t$ time steps or until the propagation fully stops.
Intuitively, we can write the relationship between the influence spread and the seed set as:
\begin{equation}\label{eq:model}
    g: \Omega \rightarrow \sigma,
\end{equation}
where $g$ is a function of seed set $\Omega$ whose size is smaller than or equal to $k$, and $g$ predicts the influence spread $\sigma$ based on the seed set $\Omega$.
To evaluate the contribution of a single seed within the set, we need to quantify the variance brought by including the seed in the seed set.

To facilitate the evaluation, an analysis of the total effect will be applied to node set. As discussed, total effect measures the amount of influence variance lost if node $i$ is not included in the seed set. It includes not only the part of the network activated by node $i$, but also the part that could be activated by this node and some other nodes at the same time.
Following Equation~\ref{eq:total_sobol}, the Sobol total index for a specific node in IM can be written as:
\begin{equation}
\small
\begin{aligned}
S^{(T)}_{i} &=\frac{\mean_{\Omega_{\sim i}}\left(\var_{\Omega_{i}}\left(Y \mid \Omega_{\sim i}\right)\right)}{\var(Y)}
    = \frac{\sum_{\Omega_{\sim i}} (Y_{\Omega_i = 1} - Y_{\Omega_i = 0})^2}{2^{k+1}\cdot \var(Y)}.
\end{aligned}
\label{eq:total_node_induction}
\end{equation}
Detailed transformation is shown in Appendix. Similar to Equation \ref{eq:short_1st_im}, Equations~\ref{eq:total_node_induction} can be rewritten as:
{\small
\begin{equation}
   S^{(T)}_{i} = \frac{\sum_{\Omega_{\sim i}} (\Delta_{\Omega_{\sim i}})^2}{2^{k+1}\cdot \var(Y)},
   \label{eq:total_node_short}
\end{equation}} 

This will be used to evaluate a specific node's global contribution to IM. Note that the numerator in Equation~\ref{eq:total_node_short} is the sum of the squares while the numerator in Equation~\ref{eq:short_1st_im} is the square of the sum.

\subsection{\algo: The Proposed Algorithm.}

Now that the Sobol total index can be used to evaluate the significance of a seed within the seed set, we can deduce that deleting the node with the lowest Sobol total index will result in the most negligible influence loss when the budget restriction is reduced from $k$ to $k-1$. Based on this finding, we propose a light-weighted plug-in approach named~\algo~as described in Algorithm~\ref{alg:1}. It can improve the performance of the mainstream IM method in two steps, namely \textbf{collecting} of seed candidates and \textbf{pruning} of the nodes with less impact on the influence spread while maintaining a good balance between effectiveness and efficiency.
\begin{algorithm}
\caption{Sobol Influence Maximization (\textbf{SIM })}\label{alg:1}
\begin{algorithmic}[1]
\REQUIRE Graph $G = (\mathcal{V}, \mathcal{E}, A)$, budget constraint $k$, number of propagation steps $t$, IM algorithm $M$
\ENSURE A $k$-sized seed set $\Omega$, i.e., $|\Omega|= k$
\STATE /* \textit{1, Collecting} */ \label{algo:over_selection_start}
\STATE Set a over-selection parameter $a\in \mathbb{R}^{+}, a > 1$
\STATE $\Omega \leftarrow M(G, \lceil ak\rceil, t)$ \label{algo:over_selection_end}
\STATE /* \textit{2, Pruning} */\label{algo:pruning_start}
\WHILE{$|\Omega| > k$}
  \FOR{seed $i$ in $\Omega$}
    \STATE Calculate $S^{(T)}_i$ \qquad (with Equation~\ref{eq:total_node_short}) \label{algo:total_sobol}
  \ENDFOR
  \STATE $i\leftarrow\arg\min_{i} S^{(T)}_i$
  \STATE $\Omega \leftarrow \Omega_{\sim i} $
 \ENDWHILE \label{algo:pruning_end}
\end{algorithmic}
\end{algorithm}

\noindent\textbf{(1) Collecting} (line \ref{algo:over_selection_start}-\ref{algo:over_selection_end}). In the first step of~\algo, $\lceil ak\rceil$ candidates are selected with state-of-the-art IM algorithms (line \ref{algo:over_selection_end}), leading to extra $\lceil ak\rceil - k$ nodes beyond the budget. IM algorithms would select overall good seed nodes ignoring higher-order interaction among them, but there may be over- or under-estimated effects, which need to be corrected.

\noindent\textbf{(2) Pruning} (line \ref{algo:pruning_start}-\ref{algo:pruning_end}). To satisfy the budget constraint, $\lceil(a-1)k\rceil$ seeds need to be removed from the $\lceil ak\rceil$ candidates. This selection process is accomplished by the Sobol total indices (line \ref{algo:total_sobol}). Note that the Sobol total index is to measure the variance lost when holding one variable stable with all other variables still in the model. Thus, removing the node with the lowest index results in the least influence loss. The pruning of the extra nodes must be done one at a time iteratively until exactly $k$ nodes are in the set. The Sobol total index for each seed left would be updated after each iteration, and the rank of the seeds might change. Thus, selecting the nodes with the largest value of indices from the candidates or removing the $\lceil ak\rceil - k$ lowest-ranking nodes at one time does not guarantee the best result. 

\noindent\textbf{Time Complexity Analysis.} For heuristic algorithms with no updates to the proxies during the selection procedure, such as degree and Eigen-centrality, the time cost of selecting $ak$ nodes is the same as selecting $k$. For those who update their proxies, the time for each node selection and proxy update iteration remains the same while the number of iterations rises from $k$ to $ak$. Thus, the time complexity of the Collecting process is at most $a*\mathcal{O}(M)$ where $\mathcal{O}(M)$ is the time complexity of $M$. This complexity is equal to $\mathcal{O}(M)$ when we consider $a$ as a constant.
Assuming that it takes $r$ rounds of simulations to estimate a seed's Sobol total index (line \ref{algo:total_sobol} in Algorithm~\ref{alg:1}), the total number of simulation rounds required for calculating the indices throughout the pruning process is $2^{ak}\cdot r + 2^{ak-1}\cdot r + \dots + 2^{k+1}\cdot r=(2^{ak+k}-2^{k+1})\cdot r =\mathcal{O}(2^{ak}r)$.
The time complexity concerning $ak$ seems exponentially huge. However, $a$ is a small constant, and $k$ is the budget restriction that does not grow with the graph size.
Besides, $r$ is a linear factor that will not expand as the size of the problem grows and will be set to an acceptable value. Therefore the extra time cost is constant regardless of the graph size as long as the budget restriction stays the same. The overall time complexity of the Algorithm~\ref{alg:1} is $a * \mathcal{O}(M) + \mathcal{O}(2^{ak}r) \sim \mathcal{O}(M)$ unless the budget constraint grows with the graph size.

\section{Experiment}


Synthetic and real-world datasets are used to test the proposed explanatory method and~\algo. 
Code for the experiments is available at 
\url{https://github.com/oates9895/SIM}

\subsection{Configurations.}
All experiments are carried out on a server with 32 AMD EPYC 7302P 16-Core processors and 32GB RAM. Simulations are performed by NDLib~\cite{rossetti2017ndlib} which is an open-source tool for investigating diffusion processes and dynamics. 

\noindent\textbf{Datasets.}
The framework is tested on seven datasets. \textbf{(1)} Five real-world datasets\footnote{Datasets at https://pytorch-geometric.readthedocs.io} including \textit{Cora, CiteSeer, PubMed, Amazon Computers, and Amazon Photo} are employed to mimic the sophisticated online social network structure. Since the IM problems traditionally focus on connected networks, the largest connected component of each graph is utilized as the network from which the seed set is selected. For Cora, CiteSeer, and PubMed, the edges are uniformly and randomly weighted between 0.40 to 0.80 for the IC model to reflect the varied activation probabilities between nodes. For Amazon Computers and Amazon Photo graphs, the edges are between 0.05 to 0.20 since those two graphs are denser, and the propagation can pass on with lower activation probabilities. For the LT model, the nodes are randomly assigned thresholds uniformly distributed between 0.01 to 0.20 in each round of simulation for each graph.
\textbf{(2)} Two synthetic graphs representing pseudo social networks are generated using NetworkX~\footnote{https://networkx.org}. The graphs include \textit{connected Watts-Strogatz small-world graphs} (SW)~\cite{watts1998collective} and {\textit{Erdős–Rényi random graphs}} (ER)~\cite{gilbert1959random}. Each graph has 5000 nodes, and the average degree is approximately 10. The edges and nodes are weighted in a similar manner to the CiteSeer graph.

\noindent\textbf{Baselines.}
To evaluate the legitimacy of our evaluation of the selected seed set and the corresponding IM method, we select a few popular IM algorithms as our candidates.~\algo~is compared with these baselines to evaluate its performance: 
\noindent\textbf{(1)} \textit{Degree Centrality (DEG)}: The first $k$ nodes with the highest degree centrality from the target graph are selected. Before each iteration, the degrees are updated such that the selected seeds are removed from the original graph. This heuristic is also known as SingleDiscount~\cite{chen2009efficient}.
\noindent\textbf{(2)} \textit{Eigenvector Centrality (EIG)}: Similarly, the first $k$ nodes are selected based on their eigenvector centrality.
\noindent\textbf{(3)} \textit{Simulation-based greedy algorithm (GRD)}~\cite{kempe2003maximizing}: In each iteration, the node with the highest marginal influence spread, which is measured by the average of 1000 rounds of simulations, is added to the seed set.
\noindent\textbf{(4)} \textit{Degree discount (DD)}~\cite{chen2009efficient}: The degree of each candidate node is discounted according to the likelihood of it being activated by the nodes that have been already selected.
\noindent\textbf{(5)} \textit{Sigma}~\cite{yan2019minimizing}: The spreading powers of the nodes are estimated by $\sum_t I\cdot A^t$ where $I$ is a unit column vector.
\noindent\textbf{(6)} \textit{Pi}~\cite{zhang2022blocking}: 
The nodes spreading powers are estimated by $I\cdot (J - \prod_{r=1} (1 - A^r))$ where $J$ is an all-one matrix and $\prod$ is the element-wise product of matrices.
Among the six, GRD is run on Cora graph only with both the IC model and the LT model to show its lack of scalability, 
DEG and EIG are applied to both the IC model and the LT model, while DD, Sigma, and Pi are only applied to the IC model.

\subsection{Results.}
The empirical study generally consists of three major parts: 
(1) \textbf{case study}: illustrate the effectiveness of the node contribution decomposition with a 5-node seed scenario. 
(2) \textbf{effectiveness verification}: demonstrate the effectiveness of the proposed algorithm against current IM algorithms.
(3) \textbf{runtime analysis}: compare the~\algo~with the baseline IM methods to evaluate its time efficiency.

\subsubsection{Case Study.}
To determine if the Sobol indices are capable of decomposing influence, we conduct a case study using a seed set created by DEG on the most linked component of CiteSeer as an illustration. Five seeds are chosen based on their degree centralities, and the IC model is utilized to estimate the Sobol indices. DEG is chosen as the sampling technique due to its simplicity and readability. And CiteSeer is chosen for computational efficiency because it is the smallest of the five graphs.
\begin{table}[htpb!]
\centering
\scalebox{0.85}{
\begin{tabular}{c|rr}
\ChangeRT{1pt}
Node Index & Sobol Total & Marginal Contribution\\ \hline
1422       & 0.305 $\pm$ 0.001     & 47.782 $\pm$ 5.960    \\
582        & 0.462 $\pm$ 0.004     & 194.512 $\pm$ 4.495        \\
1214       & 0.175 $\pm$ 0.002     & 6.916 $\pm$ 5.014        \\
2782       & 0.182 $\pm$ 0.005     & 7.220 $\pm$ 5.751       \\
1943       & 0.154 $\pm$ 0.001     & 5.354 $\pm$ 4.434        \\ \hline
\ChangeRT{1pt}
\end{tabular}}
\caption{The Legitimacy of Each Seed's Relative Contribution.}
\label{tab:1}
\end{table}
The contribution of each seed inside the set is measured by the difference between the estimated influence spreads of the node before and after its inclusion in the seed set (represented as ``marginal contribution"). The total Sobol indices for the five seeds are shown in  Table~\ref{tab:1}. The substantial positive association between a node's Sobol total index and its marginal contribution demonstrates that this index is a reliable indicator of the node's contribution. 
In addition, we analyzed the overlap amount and the distance between the nodes. When all the nodes involved are closer to one another, the influence overlaps tend to be larger because the influence flows emanating from the seeds converge early in the propagation process. If at least one set of nodes are separated by a great distance, the overlap will be small. 
As shown in Table~\ref{tab:2}, the second-order Sobol indices are proportional to the distance between any two nodes.
\begin{table}[htpb!]
\centering
\scalebox{0.82}{
\begin{tabular}{r|rrrrr}
\ChangeRT{1pt}
\multirow{2}{*}{Pairs (node index)} & 1422  & 1422   & 1422   & 1422   & 2782   \\ 
                       & 582   & 1214   & 2782   & 1943   & 1943    \\ \hline
Second-order Sobol & 0.00 & 591.60 & 619.46 & 518.00 & 340.83 \\ \cline{1-6} 
Distance               & 6     & 2      & 2      & 2      & 2     \\ \Xhline{1.5pt}

\multirow{2}{*}{Pairs (node index)} & 582  & 582  & 582  & 1214   & 1214  \\ 
                       & 1214 & 2782 & 1943 & 2782   & 1943   \\ \hline
Second-order Sobol & 0.01 & 0.00 & 0.08 & 424.81 & 329.21 \\ \cline{1-6} 
Distance               & 7    & 7    & 7    & 1      & 2      \\ \hline
\ChangeRT{1pt}
\end{tabular}
}
\caption{The second-order Sobol indices vs. distance.}
\label{tab:2}
\end{table}
\subsubsection{Effectiveness Verification.}
\begin{figure}[htpb!]
  \centering
  \includegraphics[width=0.83\linewidth]{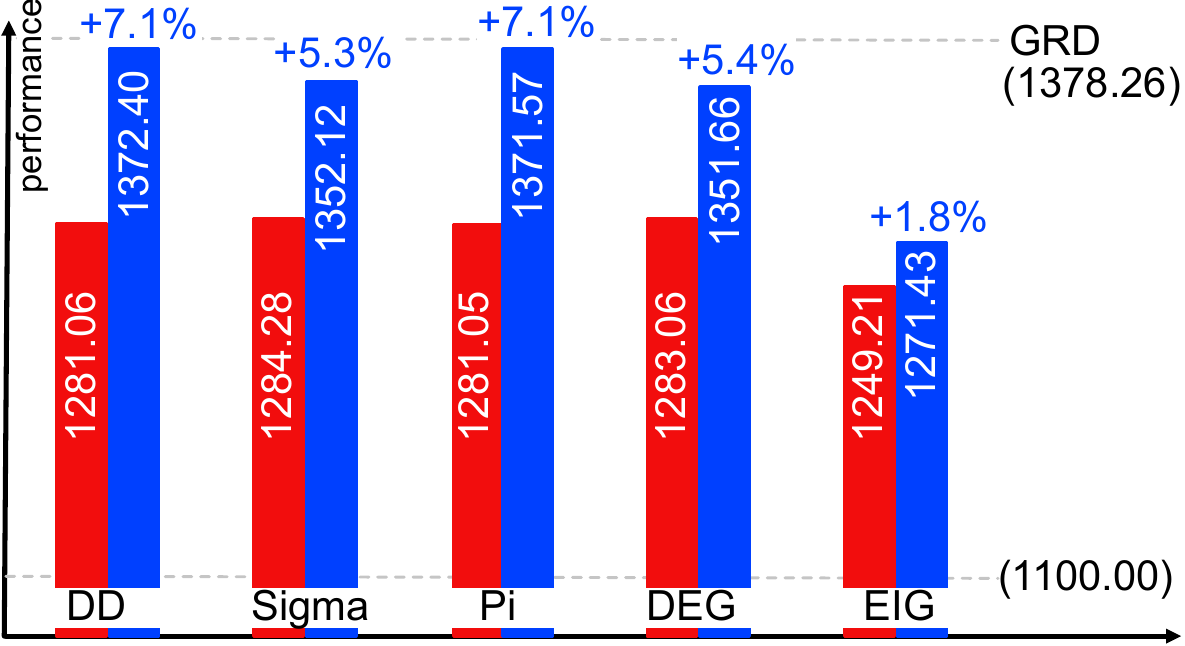}
  \caption{Performances on Cora with IC comparing to GRD.
  \textcolor{red}{Red}: Without~\algo. \textcolor{blue}{Blue}: with~\algo.}
  \label{fig:3}
\end{figure}

Under the IC model, it takes the simulation-based greedy algorithm (GRD) 17758.43 seconds to find a 5-seed set on the largest connected component of the Cora graph, who has 2485 nodes. While the running times of~\algo~and the other baselines are at most 113.62 seconds, we could observe that the expected influence spreads of the generated seed sets are almost the same. The IM performances of the six heuristics with and without~\algo~are compared with GRD in Figure~\ref{fig:3}. It shows that after combined with~\algo, the heuristics could achieve a performance almost as good as GRD. The experiment under the LT models shows a similar result. DEG with~\algo~achieves an influence spread of 1936.11 within 965.03 seconds while a similar performance of 1949.15 takes GRD 34871.65 seconds.

\begin{table*}[htpb!]
\resizebox{\textwidth}{!}{
\begin{tabular}{rr|rrrrrr}
\hline
                       &     & CiteSeer (n=2120)                    & PubMed (n=19717)                        & Computers (n=13471)                   & Photo (n=7535)                        & ER (n=5000)                           & SW (n=5000)                            \\ \hline
\multirow{2}{*}{DEG}   & W/O & 618.90 $\pm$ 34.37                   & 10948.96 $\pm$ 164.77                   & 9221.05 $\pm$ 101.55                  & 3523.68 $\pm$ 81.76                   & 4608.97 $\pm$ 68.80                   & 1284.44 $\pm$ 125.10                   \\ \cline{2-2}
                       & W   & \textbf{698.20 $\pm$ 37.62}          & 11258.95 $\pm$ 156.70                   & \textbf{9254.33 $\pm$ 118.61}         & 3581.44 $\pm$ 80.87                   & 4619.67 $\pm$ 67.84                   & \textbf{1386.88 $\pm$ 131.43}          \\ \hline
\multirow{2}{*}{EIG}   & W/O & 423.41 $\pm$ 20.34                   & 5039.46 $\pm$ 411.31                    & 9216.60 $\pm$ 105.91                  & 3456.62 $\pm$ 77.64                   & 4601.65 $\pm$ 69.25                   & 1084.60 $\pm$ 115.95                   \\ \cline{2-2}
                       & W   & 485.83 $\pm$ 24.79                   & 7639.77 $\pm$ 439.76                    & 9246.57 $\pm$ 98.83                   & 3568.40 $\pm$ 79.86                   & \textit{\textbf{4628.89 $\pm$ 63.24}} & 1227.93 $\pm$ 126.65                   \\ \hline
\multirow{2}{*}{DD}    & W/O & 672.80 $\pm$ 37.99                   & 11067.78 $\pm$ 172.69                   & 9219.99 $\pm$ 94.94                   & 3523.22 $\pm$ 78.76                   & 4593.98 $\pm$ 72.22                   & 1311.43 $\pm$ 121.18                   \\ \cline{2-2}
                       & W   & \textit{\textbf{698.79 $\pm$ 36.01}} & \textbf{11334.28 $\pm$ 178.05}          & \textit{\textbf{9263.77 $\pm$ 99.24}} & \textit{\textbf{3612.01 $\pm$ 87.54}} & \textbf{4625.59 $\pm$ 65.79}          & \textit{\textbf{1416.00 $\pm$ 117.92}} \\ \hline
\multirow{2}{*}{Sigma} & W/O & 423.82 $\pm$ 20.89                   & 11084.95 $\pm$ 162.37                   & 9214.13 $\pm$ 100.97                  & 3522.06 $\pm$ 83.07                   & 4592.14 $\pm$ 67.82                   & 1142.68 $\pm$ 105.22                   \\ \cline{2-2}
                       & W   & 694.06 $\pm$ 36.18                   & 11311.20 $\pm$ 144.72                   & 9214.29 $\pm$ 98.72                   & 3578.63 $\pm$ 82.87                   & 4613.20 $\pm$ 67.19                   & 1226.70 $\pm$ 113.71                   \\ \hline
\multirow{2}{*}{Pi}    & W/O & 619.03 $\pm$ 33.78                   & 10960.65 $\pm$ 174.90                   & 9212.89 $\pm$ 97.13                   & 3524.33 $\pm$ 81.99                   & 4597.28 $\pm$ 65.79                   & 1163.93 $\pm$ 121.87                   \\ \cline{2-2}
                       & W   & 697.63 $\pm$ 35.81                   & \textit{\textbf{11345.50 $\pm$ 151.77}} & 9251.23 $\pm$ 117.30                  & \textbf{3583.76 $\pm$ 83.45}          & 4624.04 $\pm$ 62.64                   & 1355.83 $\pm$ 120.89                   \\ \hline
\end{tabular}%
}
\caption{IM performance (i.e., influence spread) with (W) and without (W/O)~\algo~under IC model. \textbf{\textit{Italic bold}} for the best performance, \textbf{bold} for the second best.}
\label{tab:3}
\end{table*}
Figure~\ref{fig:3}~also demonstrates that applying~\algo~enhances the IM performance with a ratio up to 7\%. On the other six graphs, we compare the IM performance of the five baseline IM methods before and after they are combined with~\algo. Without~\algo, each IM algorithm generates a seed set consisting of $5$ nodes. With~\algo, each IM algorithm selects $10$ candidate nodes during the Collecting procedure and then prunes $5$ of them to meet the budget constraint. The final influence spread is measured by the mean and standard deviation of $1000$ simulations to eliminate the uncertainty. The detailed result under the IC model is presented in Table~\ref{tab:3}. See Appendix for the result under the LT model.

We observe a significant increase on influence spread after~\algo~is applied in most scenarios, with a ratio up to 60\%. The performance enhancement effect is greater when the original IM algorithm fails to generate an effective seed set. In most cases, the framework can achieve a boost of around 5\% on the influence spread. Another significant finding is that combining the degree discount heuristic with~\algo~always achieves the best or the second best performance among the algorithms.

\subsubsection{Runtime Analysis.}
It is also proved empirically that the time cost of generating the seed set increases as predicted. We observe that the increase ratios are smaller on larger graphs. This is because the time cost of pruning extra nodes is not related to the graph size, while the time cost of collecting candidate nodes scales with the graph size. Also,~\algo~runs faster on more sparse graphs since each round of simulations during the Sobol total indices estimation is faster. The details are demonstrated in Table~\ref{tab:5}~in Appendix.

\section{Conclusion}
This study presents a way of universally explaining the global effect of a seed set under the IM issue. Sobol's total index is used to assess the contribution of the seeds. The influence overlaps among seeds are quantified for the first time as an explanation using higher-order Sobol indices, which provide insight into the interaction effect and also demonstrate why the first-order method fails to accurately locate the ideal node set. 
The Sobol indices are calculated within the context of IM. Experiments indicate that the explanation we offered is applicable to graphs and IM algorithms.
A revolutionary framework, ~\algo, is provided for improving the performance of any proxy-based IM algorithm.
Experiments conducted on synthetic and real-world datasets demonstrated that our algorithm offers greater performance at an acceptable cost in terms of time.

\section*{Acknowledgement}
This work was funded by the NSF IIS award \# 2153369.

\bibliographystyle{siamplain}
\bibliography{z}
\newpage
\section*{Appendix}

\subsection*{Proof of the Equation~\ref{eq:1st_im}}

Let $a = \mean_{\Omega_{\sim i}}(Y|\Omega_i = 1)$ and $b = \mean_{\Omega_{\sim i}}(Y|\Omega_i = 0)$,
$\Gamma = \frac{1}{2}[\mean_{\Omega_{\sim i}}(Y|\Omega_i = 1) + \mean_{\Omega_{\sim i}}(Y|\Omega_i = 0)] = \frac{a+b}{2}$,

\begin{equation*}
\small
\begin{split}
S_i &= \frac{\var_{\Omega_i}(\mean_{\Omega_{\sim i}}(Y|\Omega_i))}{\var(Y)}\\
    &= \frac{\sum_{\Omega_i} (\mean_{\Omega_{\sim i}}(Y|\Omega_i)) - \Gamma)^2}{\var(Y)} \\
    &= \frac{[a - \frac{a+b}{2}]^2 + [b - \frac{a+b}{2}]^2}{2}\cdot\frac{1}{ \var(Y)} \\
    &= (\frac{a-b}{2})^2\cdot\frac{1}{\var(Y)} \\
    &=  (\frac{\mean_{\Omega_{\sim i}}(Y|\Omega_i = 1) - \mean_{\Omega_{\sim i}}(Y|\Omega_i = 0))}{2})^2\cdot\frac{1}{\var(Y)} \\
    &= (\frac{\sum_{\Omega_{\sim i}} Y_{\Omega_{i}=1} - \sum_{\Omega_{\sim i}} Y_{\Omega_{i}=0} }{2^{k-1}\cdot 2})^2\frac{1}{\var(Y)} \\
    &= \frac{1}{4^k \cdot \var(Y)}[\sum_{\Omega_{\sim i}} (Y_{\Omega_{i}=1} - Y_{\Omega_{i}=0}) ]^2
\end{split}
\end{equation*}

\subsection*{Proof of the Equation~\ref{eq:Sho}}
\begin{equation*}
\small
\begin{aligned}
    S_\Psi &= \frac{\var_\Psi  (\mean_{\Omega_{\sim \Psi}}(Y|\Psi))}{\var(Y)} \\
    &= \frac{\mean_\Psi [\mean_{\Omega_{\sim \Psi}}(Y|\Psi) - \mean_\Psi (\mean_{\Omega_{\sim \Psi}}(Y|\Psi))]^2}{\var(Y)} \\
    &= \frac{\mean_\Psi [\mean_{\Omega_{\sim \Psi}}(Y|\Psi) - \frac{1}{2^s}\sum_\Psi {\frac{\sum_{\Omega_{\sim \Psi}} (Y|\Psi)}{2^{k-s}}}]^2}{\var(Y)} \\
    &= \frac{\mean_\Psi [\mean_{\Omega_{\sim \Psi}}(Y|\Psi) - \frac{\sum Y}{2^k}]^2}{\var(Y)} \\
    &= \frac{\mean_\Psi [\mean_{\Omega_{\sim \Psi}}(Y|\Psi) - \mean(Y)]^2}{\var(Y)} \\
    &= \frac{\mean_\Psi [\frac{\Sigma_{\Omega_{\sim \Psi}}(Y|\Psi)}{2^{k-s}} - \mean(Y)]^2}{\var(Y)} \\
    &=  \frac{\sum_\Psi (\sum_{\Omega_{\sim \Psi}} (Y|\Psi) \cdot 2^{s-k} - \mean(Y))^2}{2^s\cdot\var(Y)}
\end{aligned}
\end{equation*}

\newpage
\subsection*{Proof of the Equation~\ref{eq:total_node_induction}}
\begin{equation*}
\small
\begin{aligned}
S^{(T)}_{i} &=\frac{\mean_{\Omega_{\sim i}}\left(\var_{\Omega_{i}}\left(Y \mid \Omega_{\sim i}\right)\right)}{\var(Y)} & (\text{Equation}~\ref{eq:total_sobol})\\
    &= \frac{\sum_{\Omega_{\sim i}} \var_{\Omega_i} (Y|\Omega_{\sim i})}{2^{k-1}\cdot \var(Y)} & (\text{mean of $\Omega_{\sim i}$})\\
    &= \frac{\sum_{\Omega_{\sim i}} (\mean_{\Omega_i} (Y_{\Omega_{\sim i}} - \mean(Y_{\Omega_{\sim i}}))^2}{2^{k-1}\cdot \var(Y)} \\
    &= \frac{\sum_{\Omega_{\sim i}} (\frac{Y_{\Omega_i = 1} - Y_{\Omega_i = 0}}{2})^2}{2^{k-1}\cdot \var(Y)} \\
    &= \frac{\sum_{\Omega_{\sim i}} (Y_{\Omega_i = 1} - Y_{\Omega_i = 0})^2}{4\cdot 2^{k-1} \cdot \var(Y)} & \\
    &= \frac{1}{2^{k+1}\cdot \var(Y)}\sum_{\Omega_{\sim i}} (Y_{\Omega_i = 1} - Y_{\Omega_i = 0})^2 & 
\end{aligned}
\end{equation*}

\subsection*{Uncertainty in IC and LT}

In the IC model, each active node tries to activate its neighbors with the corresponding probability $a_{ij}$ in each time step. The influence spread of a seed set has an uncertainty that comes solely from the probabilities of the nodes activating each other. This uncertainty is not what we are interested in thus we remove it by taking the average of multiple simulation rounds as the estimated total influence spread. Although evaluating the influence spread of a seed set bears a $\#P-hardness$ complexity, the  simulations carried out for the estimation of the Sobol indices are relatively feasible given $k \ll n$. Assuming the influence spread of each combination of the seeds is estimated by $r$ rounds of simulations, the total number of  simulations is $2^k r$. As a comparison, the number of simulations needed for the naive greedy IM algorithm~\cite{kempe2003maximizing} proposed by Kempe et al. is $\frac{n!}{(n - k)!}$ which is $O(n!)$ when $n \gg k$.

In the LT model, edges are weightless while each node has an individual threshold. When the percentage of a node's neighbors that are activated exceeds its threshold, the node will be activated in the next time step. In each round of the simulation, the threshold is uniformly and randomly assigned to reflect that we don't know each node's threshold.

\subsection*{Extra Experiment Results}

\begin{table*}[htpb!]
\resizebox{\textwidth}{!}{
\begin{tabular}{rr|rrrrrr}
\hline
                     &     & CiteSeer (n=2120)                     & PubMed (n=19717)                        & Computers (n=13471)                     & Photo (n=7535)                         & ER (n=5000)                            & SW (n=5000)                            \\ \hline
\multirow{2}{*}{DEG} & W/O & 933.20 $\pm$ 36.04                    & 9778.99 $\pm$ 411.22                    & 12028.66 $\pm$ 111.09                   & 4641.77 $\pm$ 107.56                   & 4069.29 $\pm$ 202.20                   & 1165.74 $\pm$ 134.95                   \\ \cline{2-2}
                     & W   & \textit{\textbf{1104.44 $\pm$ 33.46}} & \textit{\textbf{10461.96 $\pm$ 406.63}} & \textit{\textbf{12131.82 $\pm$ 114.58}} & \textit{\textbf{4792.01 $\pm$ 115.95}} & \textit{\textbf{4190.29 $\pm$ 201.99}} & \textit{\textbf{1194.85 $\pm$ 139.02}} \\ \hline
\multirow{2}{*}{EIG} & W/O & 592.14 $\pm$ 21.79                    & 3622.58 $\pm$ 204.18                    & 12022.61 $\pm$ 109.63                   & 4504.44 $\pm$ 102.02                   & 4051.96 $\pm$ 207.79                   & 891.82 $\pm$ 108.09                    \\ \cline{2-2}
                     & W   & 681.55 $\pm$ 25.22                    & 4896.88 $\pm$ 298.37                    & 12127.38 $\pm$ 110.54                   & 4739.52 $\pm$ 103.82                   & 4181.25 $\pm$ 195.91                   & 962.33 $\pm$ 119.94                    \\ \hline
\end{tabular}%
}
\caption{IM performance with (W) and without (W/O)~\algo~under the LT model.}
\label{tab:4}
\end{table*}


\begin{table}[htpb!]
\resizebox{\textwidth}{!}{%
\begin{tabular}{rr|rrrrrrr}
\hline
                       &     & CiteSeer (n=2120) & Cora (n=2485) & PubMed (n=19717) & Computers (n=13471) & Photo (n=7535) & ER (n=5000) & SW (n=5000) \\ \hline
\multirow{2}{*}{DD}    & W/O & .0.2              & 0.02          & 0.18             & 0.77                & 0.33           & 0.07        & 0.07        \\ \cline{2-2}
                       & W   & 94.15             & 108.72        & 724.74           & 2097.04             & 984.28         & 294.66      & 269.31      \\ \hline
\multirow{2}{*}{Pi}    & W/O & 1.80              & 2.43          & 147.03           & 69.05               & 22.29          & 9.73        & 9.87        \\
                       & W   & 97.30             & 113.62        & 1020.89          & 2260.50             & 1025.19        & 314.57      & 289.20      \\ \hline
\multirow{2}{*}{Sigma} & W/O & 1.33              & 1.82          & 110.75           & 52.11               & 16.63          & 7.28        & 7.31        \\ \cline{2-2}
                       & W   & 96.30             & 112.62        & 937.42           & 2222.02             & 1007.60        & 310.80      & 285.01      \\ \hline
DEG                    & W/O & 0.02              & 0.02          & 0.19             & 0.78                & 0.33           & 0.08        & 0.13        \\ \cline{2-2}
                       & W   & 94.25             & 109.29        & 665.92           & 2140.26             & 985.84         & 296.98      & 268.18      \\ \hline
\multirow{2}{*}{EIG}   & W/O & 0.10              & 0.14          & 1.55             & 4.25                & 2.03           & 0.56        & 0.65        \\ \cline{2-2}
                       & W   & 93.35             & 109.21        & 645.34           & 2143.48             & 989.88         & 298.04      & 269.19      \\ \hline
\end{tabular}%
}
\caption{Runtime comparison (in seconds)}
\label{tab:5}
\end{table}

\end{document}